\documentclass{article}
\usepackage{lineno,hyperref}

\usepackage{graphicx}
\usepackage[all]{xy}
\usepackage{amscd,amssymb,amsbsy,amsmath,amsfonts,amsthm}
\usepackage{color}
\usepackage{multicol}

\usepackage[english]{babel}
\usepackage[latin1]{inputenc}

\newcommand{\bm}[1]{\mbox{\boldmath ${#1}$}}

\numberwithin{equation}{section}

\usepackage{geometry}
\begin{document}

%TITULOS: The forgotten Term in some Path Integrals Applications

\title{Semiclassical constant elasticity of variance model for option pricing: an analytical approach }
\author{{\bf Jos\'e A.~Capit\'an}\\
\small{Departamento de Matem\'atica Aplicada y Grupo de Sistemas Complejos,}\\
\small{Universidad Polit\'ecnica de Madrid, Madrid, Spain}\\
\small{ja.capitan@upm.es}\\[2mm]
{\bf Jos\'e~Lope-Alba}\\
\small{Departamento de An\'alisis Econ\'omico: Econom{\'i}a Cuantitativa,}\\
\small{Universidad Aut\'onoma de Madrid, Madrid, Spain}\\
\small{jose.lope@uam.es}\\[2mm]
{\bf Juan J. ~Morales-Ruiz}\\
\small{Departamento de Matem\'atica Aplicada,}\\
\small{Universidad Polit\'ecnica de Madrid, Madrid, Spain}\\
\small{juan.morales-ruiz@upm.es}}

%\today
 \maketitle
 \begin{abstract}
This paper is devoted to obtaining closed-form solutions for the semiclassical approximation of the heat kernel of the diffusion equation, as defined by the constant elasticity of variance (CEV) option pricing model. This model is a natural, non-trivial extension of the classical Black-Scholes-Merton model.  One of the key points is that our calculations are based on the Van Vleck-Morette determinant instead of the Van Vleck determinant used by other authors, giving rise to highly complicated implicit formulas rather than the simple explicit formula we obtain. Another reason for this simplification is that we use the more powerful Hamiltonian formulation instead of the Lagrangian one.  In fact, we compute the Van Vleck-Morette determinant in two different ways: first, by solving the classical Hamiltonian equations, and second, by solving the variational equations. Furthermore, our calculations reveal a necessary exponential factor in the prefactor of the kernel that was missing in previous work.  
\end{abstract}

%\noindent {\bf Keywords}\keywords{ CEV option pricing model, path integrals, heat kernel propagator, semiclassical approximation, Van Vleck-Morette determinant, differential Galois theory, integrability.}

\section{Introduction: the semiclassical approximation formula}

The Pauli-Morette semiclassical approximation, also known as the Wentzel-Kramers-Brillouin (WKB) formula, which was independently introduced in 1926 by these authors in the context of quantum mechanics, is our starting point. To properly contextualize this methodology in the field of mathematical finance, we first start by illustrating how the semiclassical approximation proceeds in the context of quantum mechanics in a one-dimensional general setting. Consider a Hamiltonian system of one degree of freedom defined by the standard Hamiltonian function $H=H(x,p)=T(p)+V(x)$ and a particular classical path $\gamma$ in the configuration space, $x=x(t)$,  between the fixed points $x_1$ and $x_2$, where $t$ stands for time, $t_1$ and $t_2$ being the initial and final times, respectively. The propagator 
$$K({ x}_2,t_2 \,|\, { x}_1,t_1)$$  
is the solution of the corresponding Schr\"{o}dinger equation with initial condition $K(x_2,t_1 \,|\, { x}_1,t_1)=\delta(x_2-x_1)$.

To introduce the semiclassical approximation for the propagator, we consider the general classical solution around the
particular path by allowing variations around the initial and final positions. We define the Van Vleck-Morette determinant as the Jacobian of the final position
with respect to the initial momentum, evaluated at the final time $t_2$,

\begin{equation}J=\frac{\partial x_2}{\partial p_1}(t_2).\label{eq:jaco}\end{equation}

In fact, since the classical Hamiltonian is autonomous, $J$ depends on the elapsed time $t_2-t_1$.  The function $J$ can be obtained in two ways:

\begin{itemize}
\item[i)] Using the formula \eqref{eq:jaco} above.
\item[ii)] By means of the solution of the variational equations of the Hamiltonian system around the classical integral curve $x=x(t)$, $p=p(t)$, corresponding to the classical path $x=x(t)$ in configuration space, which is given by the linear system

\begin{equation}
\begin{pmatrix}\dot{\xi}\\ \dot{\eta}\end{pmatrix}=\begin{pmatrix}\frac{\partial^2 H}{\partial x\partial p}(x(t),p(t))&\frac{\partial^2 H}{\partial p^2}(x(t),p(t))\\
-\frac{\partial^2 H}{\partial x^2}(x(t),p(t))&-\frac{\partial^2 H}{\partial x\partial p}(x(t),p(t))\end{pmatrix}\begin{pmatrix}\xi\\ \eta\end{pmatrix}. \label{eq:VEg}
\end{equation}
Here $\xi=\delta x$ and $\eta=\delta p$ stand for the variations of position and momentum, respectively. Let $\xi=\xi(t_2-t_1,\xi_1,\eta_1)$ be the solution of the variational equation~\eqref{eq:VEg}, given the variation of the momentum at time $t_2-t_1$  with respect to the initial conditions, $\xi_1=\xi(t_1),\eta_1=\eta(t_1)$. Then $J$ will be the coefficient of $\eta_1$ in $\xi$, in $\xi$ as a linear combination of $\xi_1$ and $\eta_1$. It is clear that $J$ is the following element  of the fundamental matrix $\Phi(t_2)$ of Eq.~\eqref{eq:VEg}:

\begin{equation}
\Phi(t_2)=\begin{pmatrix}*&J\\**&*\end{pmatrix}, \label{eq:VMo}\end{equation}
with initial condition the identity matrix

$$\Phi(t_1)=\begin{pmatrix}1&0\\0&1\end{pmatrix}.$$
In other words, $J$ is given as the the variation of the position with respect to the momentum at the final time. It gives the same result as \eqref{eq:jaco} as a simple consequence of the definition of the variational equations: their solutions represent the linear part of the flow of the Hamiltonian system with respect to the initial conditions. For more details on this connection between the Van Vleck-Morette determinant and the variational equations, see \cite{MO}.

\end{itemize}

We denote by $S(\gamma)$ the action computed along $\gamma$. Then the Pauli-Morette (also known as WKB) semiclassical formula for the propagator of the Schr\"{o}dinger equation

\begin{equation} i\hbar\frac \partial{\partial t}\psi=\hat H\psi, \label{eq:Sc}\end{equation} is

\begin{equation}
K_{\textsf{WKB}}({ x}_2,t_2 \,|\, { x}_1,t_1) =A\, e^{\frac i\hbar S(\gamma)},\quad   A:=
\frac{1}{\sqrt{2\pi i\hslash J}}, \label{eq:Pauli-Morette}
\end{equation}
(see \cite{Pauli,dewitt1976semiclassical}; the formula was already implicit in Morette's early paper \cite{morette1951definition}).

This formula provides semiclassical quantum fluctuations along the particular path $\gamma$. Moreover, when the Lagrangian, or the corresponding Hamiltonian is quadratic, the semiclassical approximation is exact, and the true propagator coincides with the semiclassical one ($K=K_{\textsf{WKB}}$).
This occurs for the free particle and for the harmonic oscillator cases.

Related to the simplification that occurs when using the Van Vleck-Morette determinant \eqref{eq:jaco} in \eqref{eq:Pauli-Morette}
instead of the Van Vleck determinant,

$$\frac{\partial^2 S(\gamma)}{\partial x_1\partial x_2},$$
(they are not the same, but one is the opposite inverse of the other), Cecile DeWitt-Morette said

\medskip

``{\it ...It seems nevertheless little known by physicists (myself for example until I rediscover it) who often carry on long calculations instead of using it... }"
(\cite{dewitt1976semiclassical}, p. 371)
\medskip

Our paper will be essentially an illustration of this remark.

For more details on this formula and its connection with differential Galois theory, see \cite{MO,ALMP}. In particular, in \cite{MO}
it was proved that if the initial classical Hamiltonian system is integrable, then it is possible to obtain $K_{\textsf{WKB}}$ in closed form in the framework of differential Galois theory: the essential step in the computation is the Van Vleck-Morette determinant, and it can be done via the variational equations, whose integrability also follows from differential Galois theory. The necessary condition for the integrability of the Hamiltonian system is rigorously justified by a theorem of the third author with Ramis, see reference \cite{MR} (and also \cite{morales}).  In this paper we will only consider one degree-of-freedom Hamiltonian systems, and, as they are integrable, we can obtain the semiclassical propagator $K_{\textsf{WKB}}$ in closed form. Hence, the differential Galois theory is behind the closed-form computations presented here.

Instead of a Schr\"{o}dinger equation  \eqref{eq:Sc}, we are interested in a diffusion equation

\begin{equation} \frac \partial{\partial \tau}\psi=\hat H\psi, \, \hat H=\frac 12  q(x)\hat p^2+r(x)\hat p+V(x),
 \label{eq:diff}\end{equation}
the momentum operator, in this case, being defined as

$$ \hat p=-\frac \partial{\partial x}, $$
and $q(x)$, $r(x)$ and $V(x)$ functions of position $x$, with $q(x)\geq 0$.
We point out that the semiclassical approximation is valid not only for small $\hbar$, as in the typical semiclassical problem in quantum
mechanics,  but {\it also for small times} \cite{FEHI,KO}.

Another difference with respect to the Hamiltonian considered previously in the Pauli-Morette formula is that the classical Hamitonian,
\begin{equation}  H(x,p)=\frac 12 q(x)p^2 +r(x)p+V(x),  \label{eq:clham}\end{equation}
contains terms mixing coordinates and momenta. This fact {\it implies a modification} in the Pauli-Morette formula, by introducing an additional exponential factor in the formula. We remark that
if the Hamiltonian has the form $H=T(p)+V(x)$, then it is possible to apply the original Pauli-Morette formula, but in the so-called Euclidean time.

The extension of the Pauli-Morette formula to this class of diffusion problems is given in \cite{LRT,KO} and the Pauli-Morette formula becomes

\begin{equation}
K_{\textsf{WKB}}({ x},0 \,|\, { x}_T,T) =A\, e^{-S(\gamma)},\quad   A:=\frac{1}{\sqrt{ 2\pi J}}e^{\frac 12\int_{0}^T
 \frac {\partial^2}{\partial p\partial x} H(x(\tau), p(\tau))\, d\tau },
\label{eq:Pauli-Morettes}
\end{equation}
with initial condition
$$ K_{\textsf{WKB}}( x,T \,|\, x_T,T)=\delta(x-x_T),$$
being $(x(\tau),p(\tau))$  the classical integral curve of the Hamiltonian system in phase space with ``initial" condition $x(0)=x_T$, $p(0)=p(T):=p_T$,
where $x$ stands for the ``final" position at time $\tau=T$, i.e., $x=x(\tau,x_T,p_T)$.
Then the Van Vleck-Morette determinant is now

$$J=\frac{\partial x}{\partial p_T}(\tau=T), $$
and it will be a function of $T$ and $x_T$, $p_T$, which we can express in terms of the endpoints in configuration space, namely $x_T=x(\tau=0)$ and $x=x(\tau=T)$.
We remark that

$$\frac{\partial x}{\partial p_T}(\tau=0)=0,$$
i.e., $J=0$ for $T=0$,
because it is a non-diagonal element of the Jacobian matrix of the flow of the Hamiltonian system
$$(x_T,p_T)\mapsto (x(\tau,x_T,p_T),p(\tau,x_T,p_T)),$$
that for $\tau=0$ is the identity map.

The above notation of the initial and final points of the classical paths is motivated by the usual terminology in financial pricing problems: $T>t$ is the maturity time, and $x_T$  and $x$ are the starting and final points in the configuration space, with respect to the remaining time to maturity, $\tau=T-t$. Then the classical path connects $x_T$ and $x$ as $\tau$ increases, $0\leq\tau=T-t\leq T$.

In this contribution, we will obtain a closed-form expression for the semiclassical approximation of the propagator in diffusion problems of this kind (for diffusion problems,  the propagator is also called the heat kernel or simply the kernel). The final expression of the propagator will be given in terms of $T$ and, of course, $x_T$ and $x$.

The constant elasticity of variance (CEV) model is a well-known extension of the classical Black-Scholes (BS) framework in mathematical finance. Introduced by Cox in 1975~\cite{COX}, the CEV model addresses one of the main limitations of the BS model for derivatives pricing: its assumption of constant volatility. In the CEV model, the volatility of an asset is assumed to vary as a power function of the asset price, allowing it to increase as the price decreases ---a behavior consistent with the so-called leverage effect observed in equity markets. This feature enables the CEV model to generate more realistic volatility properties, particularly volatility skews, which are commonly observed in financial markets but not captured by lognormal models such as the BS model, which assumes a geometric Brownian motion for the underlying asset of the derivative. In addition, the CEV model provides a more realistic behavior near zero, allowing the possibility of the asset hitting zero, a feature that is not captured by the standard BS model for asset prices. 

In reference~\cite{Schroeder}, the CEV model is compared with other approaches. Its relative simplicity and ability to reflect market characteristics showcase its applicability in practical financial engineering. The CEV model has been applied to the pricing of options, interest rate derivatives, and other financial instruments, especially in contexts where the possibility of the asset reaching zero or exhibiting high volatility near lower price levels is relevant. In this contribution, we obtain closed-form formulae for the CEV option pricing model using the semiclassical approximation. Any closed-form formula that can reasonably approximate the option price in this scenario is valuable because it provides a much easier way of computing the price. In addition, since we can derive closed-form formulae, we can use them to analytically infer the effect of parameter variation on the resulting price. Although the semiclassical approximation for the CEV model was first developed by Araneda et al. (see \cite{ARVI}), the expression they obtained following a different methodology does not coincide with our result. Moreover, the WKB pricing rule obtained in that reference is much more involved than ours. Furthermore, we highlight that an exponential factor is missing in the expression derived by Araneda et al. 

We will start by illustrating the methodology summarized above with the BS model.

\section{Black-Scholes model}

\subsection{Heat kernel for the Black-Scholes model}

As a preliminary exercise, aimed at showing how the methodology proceeds, we compute the well-known propagator for the Black-Scholes model. As we will show, the classical Hamiltonian system associated to the Black-Scholes PDE is linear. Therefore, the semiclassical approximation of the BS propagator is exact, so the approximation obtained in this way will lead to the exact, classical BS pricing formula.

The BS partial differential equation (PDE) equation is (see \cite{BS})

\begin{equation}\psi_{\tau}=\frac 12 \sigma^2S^2\psi_{SS}+rS\psi_S-r\psi, \label{eq:BS}\end{equation}  written in time $\tau=T-t$,  $0\leq\tau\leq T$, for $\psi$ the price of an option given as function of the stock price $S$.

Using the well-known change of variable $e^x=S$ (where $x$ refers to the logarithm of the stock price $S$), under the risk-free assumption we obtain

\begin{equation}
\frac{\partial\psi}{\partial \tau}=\left(\frac 12 \sigma^2\frac {\partial^2}{\partial x^2}+
\mu\frac{\partial}{\partial x}-r\right)\psi=\hat H\psi, \qquad \mu:=r-\frac{\sigma^2}2. \label{eq:BSx}
\end{equation}

For this PDE, we identify the classical Hamiltonian as

\begin{equation} 
H=\frac 12\sigma^2p^2-\mu p-r.
\end{equation}
Observe that the Hamitonian is a quadratic function, hence the semiclassical approximation is exact in this case,  $K_{\textsf{WKB}}=K$.

The Hamiltonian equations are

$$\dot x=\sigma^2p-\mu, \, \dot p=0, $$
from which we trivially find

\begin{equation}
p=\frac 1{\sigma^2}(\dot x+\mu)
\end{equation}
as a first integral, being $\dot x$ also constant, i.e., the motion in $x$ is uniform

\begin{equation} 
x(\tau)=C_1+C_2\tau.
\end{equation}

From now on, the dot means derivative with respect to $\tau$.

The constant $C_1$ does not depend on the initial momentum $p(\tau=0)=p_T$. The dependence of $C_2$ on $p_T$ is

\begin{equation}
C_2=\frac{x-x_T}{T}=\dot x=\sigma^2p_T-\mu.\end{equation}
Hence, the solution of the Hamilton equations with respect to the initial conditions $(x_T,p_T)$ is
\begin{equation} 
x(\tau)=x_T+(\sigma^2p_T-\mu)\tau, \quad p=p_T, \quad 0\leq\tau\leq T, \label{eq:xtau}
\end{equation}
being the corresponding classical path $\gamma$ the projection of the above phase integral curve onto the configuration space,
\begin{equation} 
x(\tau)=x_T+\frac{x-x_T}{T}\tau, \quad 0\leq\tau\leq T. 
\end{equation}
At this point, we have obtained all the elements to compute the action on $\gamma$ and the Van Vleck-Morette determinant.

From~\eqref{eq:xtau} we find that the Lagrangian restricted to $\gamma$ is

$$
L(x(\tau))=\dot xp-H=    \frac 12\sigma^2p^2+r=\frac 1{2\sigma^2}(\frac {x-x_T}T+\mu)^2+r,
$$
which is also a first integral, i.e., it does not depend on $\tau$.  The action reduces to

\begin{equation}
S(\gamma)=\int_0^T L(x(\tau))\,d\tau=L(x(\tau))T=\frac 1{2T\sigma^2}\left(x-x_T+\mu T\right)^2+rT
\end{equation}
and the Van Vleck-Morette determinant is given by

$$
J=\frac{\partial x}{\partial p_T} (T)=\sigma^2 T.
$$
The propagator of the Black-Scholes model follows from \eqref{eq:Pauli-Morettes}:

\begin{equation}\label{eq:KBS}
K= \frac{1}{\sqrt{ 2\pi J(T)}}\, e^{-S(\gamma)}=\frac{1}{\sqrt{ 2\pi \sigma^2 T}}\,e^{-\frac 1{2T\sigma^2}\left(x_T-x-\mu T\right)^2-rT}.
\end{equation}
Recall that, as the classical Hamiltonian system is linear, the semiclassical approximation of the propagator is exact. Note also that the term
$$ 
e^{\frac 12\int_{0}^T
\frac {\partial^2}{\partial p\partial x} H(x(\tau), p(\tau))\, d\tau }
$$
in this example is equal to 1, so it does not introduce any additional term on the expression because the non-dual dependence in the Hamiltonian of the moments and coordinates holds in this case.

\subsection{Pricing function of an European call option under the Black-Scholes  kernel}

The solution of the Back-Scholes equation~\eqref{eq:BSx} amounts to applying the propagator to the initial condition of the PDE. For an European call option, the price reduces to

\begin{equation} \label{eq:psi}
\psi(x,0)=\frac{e^{-rT}}{\sqrt{2\pi\sigma^{2}T}}\int_{-\infty}^{\infty}e^{\frac{-1}{2\sigma^{2}T}(x_{T}-x-\mu T)^{2}}\psi_{T}(x_{T})dx_{T},
\end{equation}
where, at maturity ($\tau = T$), the price of the option must be $\psi_{T}(x_{T})=\max\{e^{x_T}-E,0\}$, being $E$ the strike price of the call option. Simple manipulations yield

\begin{equation} \label{eq:I1}
\psi(x,0)=\frac{e^{-rT}}{\sqrt{2\pi\sigma^{2}T}}\int_{\log E}^{\infty} e^{\frac{-1}{2\sigma^{2}T}(x_{T}-x-\mu T)^{2}}(e^{x_{T}}-E)dx_{T},
\end{equation}
which is equal to the convolution between the propagator and the boundary condition for the underlying asset:
\begin{equation} \label{eq:Kpsi}
\psi(x,0)=K*\psi_{T}(x_{T})
\end{equation}

Changing back to the original variable $S=e^{x}$, doing some simplifications and regrouping in a well-known manner to obtain the error function, we get the cumulative distribution function $N(\cdot)$ of the standard normal distribution. It is left to the reader to follow \cite{BS} to obtain the contract function:

\begin{equation} \label{eq:SOL}
\psi(S,0)=S_{0}N(d_{1})-Ee^{-rT}N(d_{2}),
\end{equation}
where $d_{1}=\frac{\log(\frac{S_{T}}{E})-\tau(r+\sigma^{2})}{\sqrt{\sigma^{2}T}}$ and $d_{2}=d_{1}-\sigma\sqrt{T}$, which are basically the two probabilities of the final state at maturity $T$ of the underlying price $S$ with respect to the strike price $E$.

\section{CEV model}

\subsection{The backward Kolmogorov equation for the CEV model}

In 1975, Cox \cite{COX} introduced the so-called constant elasticity of variance model, which is defined as follows.

Given a continuous-time random process, $X_t$, defined by the stochastic differential equation

\begin{equation}
dX_t=\mu(X_t)dt+\sigma(X_t)dW_t, \label{eq:rand}
\end{equation}
where $W_t$ stands for a Brownian motion, and $\mu(x)$ and $\sigma(x)$ are suitable functions of $x$. Then to \eqref{eq:rand} it is associated with a diffusion equation, the so-called backward Kolmogorov equation.

\begin{equation}
\frac{\partial}{\partial t}\varphi=\frac 12 \sigma^2(x)\frac{\partial^2}{\partial x^2}\varphi+\mu(x)\frac{\partial}{\partial x}\varphi, \label{eq:bK}
\end{equation}
where $\varphi(x,t)$ stands for the probability that the random process takes the value $x$ at time $t$. Because we work with the backward Kolmogorov equation in this case, we do not include the discount rate $r$ for the option payoff in the drift term of the stochastic process, as we did before while discussing the Black-Scholes equation; cf. Eq.~\eqref{eq:BSx}. Therefore, when we obtain the option pricing formula after the calculation of the propagator, we have to account for the discount rate by multiplying the result by the factor $e^{-rT}$ (see section~\ref{sec:3.3}). Observe that this factor naturally arises in the calculation for the BS model (see Eq.~\eqref{eq:KBS} applied to the initial condition $\max\{S_T-E,0\}$).
 
The CEV model is defined by assuming the following stochastic process for the stock price $S$:

\begin{equation}
dS_t=\mu S_t dt+\sigma S_t^{\alpha+1}dW_t, \label{eq:randCEV}
\end{equation}
namely $\mu$, $\sigma$, and $\alpha$ constants. Under the assumption of risk-neutral markets, the drift parameter must be chosen as $\mu=r$, $r$ being the interest rate, as in the standard BS model. In addition, if the stock is intended to pay dividends by an amount $d$, the drift parameter $\mu$ has to be replaced by $r-d$~\cite{Linetsky}. For the sake of generality, we will carry out the analysis for arbitrary $\mu$ in the rest of the contribution.

Following \cite{COX}, we assume that $-1\leq \alpha<0$. In reference \cite{COX} it is used $\beta=2(\alpha+1)$ instead of our exponent $\alpha$. For $\alpha=0$, the CEV model reduces to the Black-Scholes model, although in a singular way (as evidenced by the change of variables introduced in the next step). Later, it will be relevant that $\alpha<0$.

By means of a change of variable, the CEV model reduces to one studied by Feller in 1951 \cite{FE,CORO}.   Assuming $\alpha\neq 0$, let us define the new random process $X_t$ from $S_t$ from

$$ S_t^{-2\alpha}=\sigma^2\alpha^2X_t.$$
Then the following stochastic differential equation is obtained for $X_t$:

$$dX_t=(2+\frac 1\alpha-2\alpha\mu X_t)dt-2\sqrt{X_t}dW_t,$$
The associated backward Kolmogorov equation reduces to

\begin{equation}
\frac{\partial}{\partial \tau}\psi= \left(2x\frac{\partial^2}{\partial x^2}+(a-bx)\frac{\partial}{\partial x}\right)\psi, \qquad a=2+\frac 1\alpha,\, b=2\alpha\mu.
\label{eq:fpCEV}
\end{equation}
Equation \eqref{eq:fpCEV} is of the type studied by Feller (see \cite{FE}),

\begin{equation}
\frac{\partial\psi}{\partial \tau}= \frac{\partial^2}{\partial x^2}(\beta x\psi)-\frac{\partial}{\partial x}[(\gamma+\delta x)\psi], \label{eq:fell}
\end{equation}
$\beta>0$, $\gamma$ and $\delta$ being constant parameters. The reduction of the CEV model to the diffusion equation studied by Feller in \cite{FE} was pointed out for the first time by Cox \cite{COX}.
 
We begin with the diffusion PDE given by Eq.~\eqref{eq:fpCEV}. By applying a Laplace transform in time, the Cauchy problem for \eqref{eq:fpCEV} can, in principle, be analyzed in terms of confluent hypergeometric functions. This suggests that one might solve \eqref{eq:fpCEV} by inverting the Laplace transform. However, a fundamental obstacle arises: confluent hypergeometric equations are not integrable in the sense of differential Galois theory ---that is, they are not solvable by quadratures in any reasonable sense. Consequently, there is no hope of obtaining a closed-form solution to the Cauchy problem for \eqref{eq:fpCEV}. We leave the technical details of this analysis to Appendix~\ref{sec:app2}.

Nonetheless, as pointed out in the introduction, the corresponding classical Hamiltonian system has one degree of freedom and is therefore integrable. This makes it possible to obtain a semiclassical approximation to the solution in terms of closed-form analytical expressions. This observation motivates the main goal of our paper: {\it to derive a semiclassical closed-form expression for the heat kernel of equation~\eqref{eq:fpCEV}}. Although this problem was previously studied in \cite{ARVI}, our treatment considerably simplifies the analysis.
 
%We begin with the diffusion  PDE given by Eq.~\eqref{eq:fpCEV}. By means of  a Laplace transform in time, it is possible to study the Cauchy problem of \eqref{eq:fpCEV}  by confluent hypergeometric functions. Then in principle, it will possible to solve \eqref{eq:fpCEV} by inverting the Laplace transform; but the problem is that  {\it generically} the confluent hypergeometric equations are not integrable in the sense of the differential Galois theory, that is they are not solvable by quadratures in any reasonable sense, and hence it will  no chance  to solve the Cauchy problem for  equation \eqref{eq:fpCEV} in closed form:  we left the details  in  Appendix \ref{sec:app2}.  However, as it was pointed out in the introduction, as the classical Hamiltonian system is of one  degree of freedom, an hence integrable, it will possible to obtain the semiclassical approximation of the solution by closed analytical formulas. This motivated the main goal of our paper as  {\it  to obtain a semiclassical closed-form solution for the heat kernel of \eqref{eq:fpCEV}}. This problem was studied in \cite{ARVI}, but  here we simplify considerably the calculations.
 
%the Green function of \eqref{eq:fell} as a confluent hypergeometric function, but we will not follow this approach here. {\it Our objective is to obtain a semiclassical closed-form solution for the heat kernel of \eqref{eq:fpCEV}. } This problem was studied in \cite{ARVI}, but  here we simplify the calculations.

\subsection{Heat kernel for the CEV model}

As with the Black-Scholes model, we start solving the classical Hamiltonian system associated to the diffusion equation \eqref{eq:fpCEV}. The classical Hamiltonian is

\begin{equation} H=2x p^2+(bx-a) p. \label{eq:clH}\end{equation}
As the Hamiltonian is cubic, the semiclassical approximation is not exact.

The Hamilton equations are

\begin{equation}\dot x=4xp+bx-a, \quad \dot p=-2p^2-bp, \label{eq:HCEV}\end{equation}
with initial conditions

$$x(\tau=0)=x_T,\quad p(\tau=0)=p_T,$$
being $x(\tau=T)=x$. We recall that the dot means derivative with respect to $\tau$.

We first solve the system for the momentum variable:

\begin{equation}
p(\tau)=\frac b{C_1e^{b\tau}-2}. \label{eq:mosol}
\end{equation}
Observe that, in order to match the initial condition for the momentum, it must hold that

\begin{equation}C_1=\frac b{p_T}+2.\label{eq:C1pT}\end{equation}
Then
\begin{equation}
p(\tau)=\frac {bp_T}{(b+2p_T)e^{b\tau}-2p_T}. \label{eq:mosol2}
\end{equation}
Now, substituting \eqref{eq:mosol} into the first equation of \eqref{eq:HCEV}, the equation for the position becomes an elementary linear differential equation with solution

\begin{equation}
x(\tau)=d-4C_1C_2+C_1^2C_2 e^{b\tau}+(4C_2-\frac{2d}{C_1})e^{-b\tau},\quad d:=\frac ab, \label{eq:posol}
\end{equation}
where we have defined the new parameter $d=a/b$,  and $C_2$ stands for the second integration constant.

\begin{equation}C_2 = -\frac{p_{T}^2 \left( (a - x_{T})b - 2p_{T}x_{T} \right)}{b^2 (b + 2p_{T})}.\label{eq:C2pT}\end{equation}
 
We note that, by construction, the constants $C_1$ and $C_2$ are positive because they are given by suitable exponential functions.

By recasting the constants, it is possible to write \eqref{eq:posol} as
\begin{equation} x(\tau)=D_1+D_2e^{b\tau}+\frac{D_1^2-d^2}{4D_2}e^{-b\tau}=\frac{(D_1+2D_2e^{b\tau})^2-d^2}{4D_2e^{b\tau}}, \label{eq:posol2}
\end{equation}
being

\begin{equation}
D_1=d-4C_1C_2,\quad  D_2=C_1^2C_2.  \end{equation}
Equivalently,

\begin{equation} C_1=\frac{4D_2}{d-D_1}, \, C_2=\frac{(d-D_1)^2}{16D_2}.  \label{eq:const}
\end{equation}

Observe that Eq.~\eqref{eq:posol2} coincides with the solution obtained in \cite{ARVI} using the Lagrange equations.
Hence, the solution of the Hamiltonian system is

\begin{equation} x= \frac{(D_1+2D_2e^{b\tau})^2-d^2}{4D_2e^{b\tau}}, \quad p= \frac{b}{C_1e^{b\tau}-2}= \frac{b(d-D_1)}{4D_2e^{b\tau}+2D_1-2d}. \label{eq:CIDs}\end{equation}

Now, the solution above can be written as a function of the initial conditions $x_T=x(\tau=0)$ and $p_T=p(\tau=0)$ by substituting the constants $D_1$ and $D_2$ by
\begin{equation}
D_1=-\frac{8x_T}{b^2}p_T^2-\frac{4(x_T-d)}b p_T+d, \quad D_2=\frac{4x_T}{b^2}p_T^2+\frac{2(2x_T-d)}{b}p_T+x_T-d. \label{eq:HsolCI}
\end{equation}
So,

\begin{equation} 
\begin{split}  
x = \frac{1}{b^2}[d \,b^{2}+4 \left(d-{x_T}\right) {p_T} b-8 {p_T}^{2} {x_T}+
\left(2p_T+b\right) \left(2 {x_T} {p_T}+x_Tb-db\right){\mathrm e}^{b \tau}+\\\left(4 {p_T}^{2} {x_T}-2 {p_T} b d\right)
{ e}^{-b \tau}]. 
\end{split} \label{eq:Xptxt}
\end{equation}

From \eqref{eq:HCEV}, the momentum can also be written as a function of the position $x$ and its time derivative $\dot x$ as:
\begin{equation}
p=\frac{\dot x+bd-bx}{4x}.\label{eq:mo3} 
\end{equation}
At this point, we have all the elements for computing the semiclassical approximation of the propagator.

First, we compute the integration constants $D_1$ and $D_2$ with respect to the initial and final points of the path in the configuration space, $x(0)=x_T$ and $x(T)=x$. From \eqref{eq:posol2} it follows the system of equations

$$ 
x_T=\frac{(D_1+2D_2)^2-d^2}{4D_2}, \quad x=\frac{(D_1+2D_2e^{bT})^2-d^2}{4D_2e^{bT}}, 
$$
which solved for the integration constants $D_1$ and $D_2$ leads to the following expressions in terms of $x$ and $x_T$,

\begin{equation}
D_1=  \frac{\left({\mathrm e}^{\mathit{bT}}+1\right) \sqrt{d^{2} \left({\mathrm e}^{b T}-1\right)^{2}+
4 x \mathit{x_T} {\mathrm e}^{b T}}-2\left(x+\mathit{x_T}\right) {\mathrm e}^{\mathit{bT}} }{\left({\mathrm e}^{b T}-1\right)^{2}},\label{eq:D1}
\end{equation}

\begin{equation}
D_2=\frac{x {\mathrm e}^{\mathit{bT}}+\mathit{x_T}-\sqrt{d^{2} \left({\mathrm e}^{b T}-1\right)^{2}+4 x \mathit{x_T}
 {\mathrm e}^{b T}}}{\left({\mathrm e}^{b T}-1\right)^{2}},\label{eq:D2}
\end{equation}
which coincide with the expressions obtained in \cite{ARVI}.

Now we have to compute: the action $S(\gamma)$, the integral

$$
\int_0^T \frac{\partial^2 H}{\partial x\partial p}(\tau)\, d\tau,
$$
and the Van Vleck-Morette determinant $J$.

In order to evaluate the action, we first obtain the Lagrangian as a function of time $\tau $, which turns out to be

$$
L=2p^2x=\frac {b^{2} \left(d-{D_1}\right)^{2} \left[\left(d+{D_1}\right) { e}^{-b\tau}+
2 {D_2}\right]}{8 {D_2} \left(2 {D_2} { e}^{b \tau}+{D_1}-d\right)}.
$$
Thus, the action on the classical path is given by
\begin{multline}
S(\gamma)=\int_0^T\frac {b^{2} \left(d-{D_1}\right)^{2} \left[\left(d+{D_1}\right) { e}^{-b\tau}+
2 {D_2}\right]}{8 {D_2} \left(2 {D_2} { e}^{b \tau}+{D_1}-d\right)}\,d\tau=\\
\left[\frac{ db}2\log\! \left(2{D_2} { e}^{b \tau}+{D_1}-d\right) +\frac{b}{8D_2}(d^2-D_1^2) { e}^{-b \tau}-\frac 12 db^2\tau\right]_{\tau=0}^{\tau=T}
\end{multline}
which, after simplifications, results in
\begin{equation}
S(\gamma)=\log\left[\left(\frac{2{D_2} { e}^{b T}+{D_1}-d}{2D_2+D_1-d}\right)^{bd/2}\right] +\frac{b}{8D_2}(d^2-D_1^2) ({ e}^{-b T}-1)-\frac 12 db^2T.
\label{eq:SCEV}
\end{equation}

Secondly,

$$
\frac{\partial^2 H}{\partial x\partial p}=4p+b=-b+\frac{4 {D_2} b { e}^{b \tau}}{2 {D_2} { e}^{b \tau}+{D_1}-d}.
$$
Integrating over time, we find

\begin{equation}
\int_0^T\frac{\partial^2 H}{\partial x\partial p}\,d\tau=-bT+2\log\left(\frac{2D_2e ^{bT}+D_1-d}{2D_2+D_1-d}\right). \label{eq:expA}
\end{equation}

It remains to obtain the van Vleck-Morette determinant. From equation \eqref{eq:posol2}, to compute

$$
J=\frac{\partial x}{\partial p_T}(T)
$$
we have to obtain the derivatives of the constants with respect to the momentum. This calculation yields
$$
\begin{aligned}
&\frac{\partial D_1}{\partial p_T}(T)=-\frac{16x_T}{b^2}p_T+\frac{4(d-x_T)}{b}=\frac{D_1^2-4D_2^2-d^2}{bD_2},\\
&\frac{\partial D_2}{\partial p_T}(T)=\frac{8x_T}{b^2}p_T+\frac{2(2x_T-d)}{b}=\frac{4D_2+2D_1}{b},
\end{aligned}
$$
as follows from \eqref{eq:D1}, \eqref{eq:D2} and \eqref{eq:CIDs}. Hence, the Van Vleck-Morette determinant becomes
\begin{equation}
J=\frac 1{bD_2}\left[D_1^2-4D_2^2-d^2+(4D_2^2+2D_1D_2)e^{bT}+(d^2-2D_1D_2-D_1^2)e^{-bT}\right].\label{eq:JJ}
\end{equation}
As a second verification, in Appendix~\ref{sec:app} we compute the Van Vleck-Morette determinant \eqref{eq:JJ} by means of the variational equation. We observe that for $T=0$, $J=0$ as it should be. Moreover, from the formula \eqref{eq:Xptxt} another expression of $J$ is obtained as a function of the initial conditions in phase space.
\begin{equation}
J=\frac{4 db-4 \mathit{x_T} b-16 \mathit{x_T} \mathit{p_T}+\left(4 \mathit{x_T}b+8 \mathit{x_T} \mathit{p_T}-2db\right) {\mathrm e}^{b T}+\left(8 \mathit{x_T} \mathit{p_T}-2db\right) {\mathrm e}^{-b T}}{b^{2}} \label{eq:Jphase}. 
\end{equation}
Then as 
$$\frac{\partial J}{\partial T}(T=0)=4x_T>0,$$ 
necessarily $J(T)>0$, for a sufficiently small $T>0$, as it should be.  

As a result of this calculation, we can write the semi-classical propagator $K$ in terms of the Pauli-Morette formula \eqref{eq:Pauli-Morettes} with the action given by Eq. \eqref{eq:SCEV}, and the prefactor being explicitly determined by Eqs. \eqref{eq:expA} and \eqref{eq:JJ}:
\begin{align}
K_{\textsf{WKB}} &=
\frac{
    e^{
        -\frac{1}{2}bT + \log
            \left(\frac{2D_2 e^{bT} + D_1 - d}{2D_2 + D_1 - d}\right)
    }
}{
    \sqrt{
            \frac{2\pi}{bD_2} \left[
                D_1^2 - 4D_2^2 - d^2
                + (4D_2^2 + 2D_1D_2)e^{bT}
                + (d^2 - 2D_1D_2 - D_1^2)e^{-bT}
            \right]
    }
} \times \nonumber \\
&\quad e^{-
    \frac{bd}{2}\log
        \left(\frac{2D_2 e^{bT} + D_1 - d}{2D_2 + D_1 - d}\right)
    + \frac{b}{8D_2}(D1^2 - d^2)(e^{-bT} - 1)
    + \frac{1}{2} db^2T
}. \label{eq:Kp_CEV1}
\end{align}
After some simplifications, we obtain
\begin{equation}
K_{\textsf{WKB}} =
\frac{
    \left(\frac{2D_2 e^{bT} + D_1 - d}{2D_2 + D_1 - d}\right)^{1-\frac{bd}{2}}e^{\frac{1}{2}bT(db-1)+
     \frac{b}{8D_2}(D_1^2 - b^2)(e^{-bT} - 1)
}}
{
    \sqrt{
            \frac{2\pi}{bD_2} \left[
                D_1^2 - 4D_2^2 - d^2
                + (4D_2^2 + 2D_1D_2)e^{bT}
                + (d^2 - 2D_1D_2 - D_1^2)e^{-bT}
            \right]
    }
} \label{eq:Kp_CEV}
\end{equation}
with the constants $D_1$ and $D_2$ are to be expressed as functions of $x_T$ and $x$ (see formulas \eqref{eq:D1} and \eqref{eq:D2}, respectively).

The reader can compare our result for the propagator, cf. Eq.~\eqref{eq:Kp_CEV}, with the one obtained by Araneda et al. in reference \cite{ARVI}. Importantly, the exponential factor

$$ 
e^{\frac 12\int_{0}^T
\frac {\partial^2}{\partial p\partial x} H(x(\tau), p(\tau))\, d\tau }
$$
was missing in that reference \cite{ARVI}.

\subsection{Pricing function of an European call option under the CEV kernel}
\label{sec:3.3}

At maturity, the payoff of the resulting option is $\max\{S_T - E, 0\}$, $E$ being the strike price. The payoff at maturity must be discounted by a constant rate $r$. After the change of variables
$$
S_T^{-2\alpha} = \sigma^2 \alpha^2 X_T,
$$
the payoff at maturity becomes
$$
\psi_T(x_T) = e^{-rT} \max\left\{\frac{1}{(\sigma^2 \alpha^2 x_T)^{1/(2\alpha)}} - E, 0\right\}.
$$
The payoff $\max\left\{\frac{1}{(\sigma^2 \alpha^2 x_T)^{1/(2\alpha)}} - E, 0\right\}$ is non-zero when
$$
\frac{1}{(\sigma^2 \alpha^2 x_T)^{1/(2\alpha)}} \geq E,
$$
which after some elementary algebra reduces to
$$
x_T \geq \frac{E^{-2\alpha}}{\sigma^2 \alpha^2},
$$
because $\alpha<0$.
Therefore, applying the semiclassical approximation for the propagator, we obtain the pricing function
\begin{equation} \label{eq:Kpsi_CEV1}
\psi(x,0)=K_{\textsf{WKB}} * \psi_{T}(x_{T}),
\end{equation}
which reduces to
\begin{equation}\label{eq:psi_CEV2}
\psi(x,0) = e^{-rT}\int_{\left(\sigma\alpha E^\alpha \right)^{-2}}^{\infty}
K_{\textsf{WKB}}(x,0\vert x_T,T) ((\sigma^2 \alpha^2 x_T)^{-1/(2\alpha)} - E)\, dx_T.
\end{equation}
Finally, as $\alpha = \frac{1}{a-2}= \frac{1}{bd-2}$, then

\begin{equation}\label{eq:psi_CEV}
\psi(x,0) = e^{-rT}\int_{\left(\frac{2-bd}{\sigma}\right)^2 E^{\frac{2}{2-bd}}}^{\infty}
K_{\textsf{WKB}}(x,0\vert x_T,T)\left[ \left (\frac{\sigma}{2-bd}\sqrt{x_T}\right)^{2-bd}-  E\right]\, dx_T.
\end{equation}
The contract function $\psi(x,0)$ has been written in terms of the parameter combinations that arise in the formula for the semiclassical approximation of the propagator, namely $b$ and $d$.

In order to test the validity of the closed-form formula obtained for the CEV model, we have carried out Monte Carlo simulations of the model and evaluated the degree of accuracy obtained by the formulae. These results are summarized in Appendix~\ref{sec:app3}. Simulations confirm that the semiclassical approximation performs well for low maturities, and when the exponent $\alpha$ takes values from $-0.4$ to $-1$ (see Fig.~\ref{fig:alpha}), leading to more skewed volatility values. This has important implications in finance because the CEV model was conceived to account for tail risk and volatility skew, and the semiclassical approximations can reproduce option prices very well for large values of the skew. In contrast, the error tends to increase as the maturity time increases or as the parameter $\alpha$ approaches zero, which is associated with smaller skew and tail risk effects, as in the BS model. This aligns with theoretical expectations regarding the limits of the semiclassical approximation and highlights the suitability of using the WKB formulae for option pricing in more extreme market environments.

\bigskip

\subsection*{Acknowledgements.}

JAC acknowledges financial support from grant PRIORITY (PID2021-127202NB-C22), funded by Agencia Estatal de Investigaci\'on (MCIN/AEI/10.13039/501100011033) and ``ERDF. A way of making Europe''. JJMR has been supported by Universidad Polit\'ecnica de Madrid research group \emph{Modelos Matem\'aticos no lineales}.

\bigskip

\renewcommand{\thefigure}{\thesection.\arabic{figure}}

\appendix

\section{Reduction of the CEV model to the confluent hypergeometric equation}
\label{sec:app2} 

In this Appendix, we highlight the challenges involved in solving the Cauchy problem for equation \eqref{eq:fpCEV} using the Laplace transform.

The Laplace transform of \eqref{eq:fpCEV} with respect to time is 
\begin{equation}  \left(2x\frac{d^2}{d x^2}+(a-bx)\frac{d}{d x}\right)y=sy+\psi_0(x), \label{eq:CEVlap}\end{equation} 
where we denote $y=\hat{\psi}(x,s)$ the Laplace transform of $\psi(x,\tau)$ with respect to time. The initial condition for equation \eqref{eq:fpCEV} is $\psi(x,0)=\psi_0(x)$. 

By making the change 

$$x=\frac 2b z,$$
 the homogeneous part of equation \eqref{eq:CEVlap} 
 
 \begin{equation}  \left(2x\frac{d^2}{d x^2}+(a-bx)\frac{d}{d x}-s\right)y=0, \label{eq:CEVlap1}\end{equation} 
transforms into the Kummer's differential equation
 
 \begin{equation}  \left(z\frac{d^2}{d z^2}+(\nu-z)\frac{d}{d z}-\chi\right)y=0, \label{eq:kum}\end{equation} 
 with parameters 
 
 $$\chi=\frac sb,\,\, \nu=\frac a2.$$
 
The Kummer's equation \eqref{eq:kum} is one of the standard forms of the confluent hypergeometric equations. The characterization of the integrability or solvability by quadratures of \eqref{eq:kum} was obtained by Martinet and Ramis in the framework of the differential Galois theory as follows (see \cite{MARA}, Proposition 3.1.8). Equation \eqref{eq:kum} is integrable if and only if either one of the conditions is satisfied:
 
\begin{itemize}
\item[i)] Either  $\chi$ or $\chi-\nu+1$ is a negative integer, or
\item[ii)] Either $1-\chi $ or $\nu-\chi$ is a negative integer.
\end{itemize}

As any of the above conditions for integrability are not satisfied for an arbitrary real value of the Laplace parameter $s$, then any attempt to obtain closed analytical formulas for the Cauchy problem of \eqref{eq:fpCEV} will fail. For more information on the concept of integrability in the framework of the differential Galois theory, see \cite{morales}, and for a similar obstruction to obtain closed-form solutions by means of the Laplace transform for a different diffusion equation, see \cite{ACM}.

\bigskip

\section{The variational equations for the CEV model}
\label{sec:app}

In this Appendix, we illustrate the method of computation of the Van Vleck.Morette determinant by means of the solution of the variational equation \eqref{eq:VMo}. We use the time to maturity $\tau$, i.e., the initial conditions are $\xi_0$, $\eta_0$, and the final time $\tau=T$.

The Hamiltonian is \eqref{eq:clH} and the variational equation \eqref{eq:VEg} becomes

\begin{equation}
\begin{pmatrix}\dot{\xi}\\ \dot{\eta}\end{pmatrix}=\begin{pmatrix}4p+b&4x\\
0&-4p-b\end{pmatrix}\begin{pmatrix}\xi\\ \eta\end{pmatrix}, \label{eq:VECEV}
\end{equation}
being $x=x(\tau)$ and $p=p(\tau)$ the classical solution of the Hamiltonian system, \eqref{eq:posol} and \eqref{eq:mosol}. We remark that now this particular solution of the Hamiltonian system is fixed, i.e., {\it the integration constants $D_1$ and $D_2$ are fixed.}

As the classical Hamiltonian system is integrable, because it is a 1-degree of freedom, we know by a theorem of the third author with Ramis (see \cite{MO} and references therein)  that the variational equation must also be solved in closed form, in the sense of the differential Galois theory.

We start by solving the variation of the momentum

\begin{equation}
\eta \! \left(\tau\right)=\frac{{A_1} { e}^{b \tau}}{\left(2 {D_2} { e}^{b \tau}+{D_1}-d\right)^{2}}, \label{eq:tau}
\end{equation}
being $A_1$ the integration constant. Using the last expression, we can solve the variation of the position by means of variation of constants:

\begin{equation}
\xi(\tau) = \frac{1}{2bD_2^2} \big[
    (2bA_2(d-D_1)^2D_2^2 - A_1D_1)e^{-b\tau} + 8bA_2D_2^4e^{b\tau}
    - 8bA_2(d-D_1)D_2^3 - 2A_1D_2
\big], \label{eq:xitau}
\end{equation}
being $A_2$ the other integration constant.  The equations \eqref{eq:tau} and \eqref{eq:xitau} are the solutions of the variational equations, but we need to write them as a function of the initial conditions.

The integration constants depend on the initial conditions in time $\tau$, $\xi_0=\xi(\tau=0)$ and $\eta_0=\eta(\tau=0)$, as the linear combination

\begin{equation} 
A_1=(2D_2+D_1-d)^2\eta_0\quad A_2=\frac{2D_2+D_1}{2bD_2^2}\eta_0+\alpha\xi_0,
\end{equation}
where we will not need the coefficient of $\xi_0$: only the  dependence with respect to $\eta_0$ will be relevant here.

By substituting the above equations in \eqref{eq:xitau}, for $\tau=T$, we obtain the Van Vleck-Morette determinant as  the coefficient of $\eta_0$, i.e., $\xi  \left(T\right)=J\eta_0+\cdots $, which turns out to be
$$
J=\frac 1{bD_2}\left[D_1^2-4D_2^2-d^2+(4D_2^2+2D_1D_2)e^{bT}+(d^2-2D_1D_2-D_1^2)e^{-bT}\right],
$$
the same result obtained in Eq. \eqref{eq:JJ}.

\bigskip

\section{Numerical validation of the semiclassical approximation for the CEV model}
\label{sec:app3}

This Appendix is devoted to illustrate the range of validity of the semiclassical approximation formulae obtained for the CEV model. For that purpose, we generated Monte Carlo model simulations for variable key parameters, such as the maturity time $T$ and the exponent $\alpha$ controlling volatility skew. 

Monte Carlo simulations proceed as follows: we first generate multiple asset price paths over time by discretizing the CEV stochastic differential equation~\eqref{eq:randCEV} following the Euler-Maruyama method, using antithetic variates applied to reduce simulation variance~\cite{glasserman}. Then, given the price of the stock at maturity $S_T$, we compute the value of the European call option as the expected discounted payoff $e^{-rT}\max\{S_T - E, 0\}$ from the simulated asset terminal price, $E$ being the strike value. This approach is grounded in standard quantitative finance techniques, as discussed in~\cite{glasserman}. We then measured the error between model simulations and the approximation given by the WKB formula.

\begin{figure}[t!]
\centering
\includegraphics[width=0.65\textwidth]{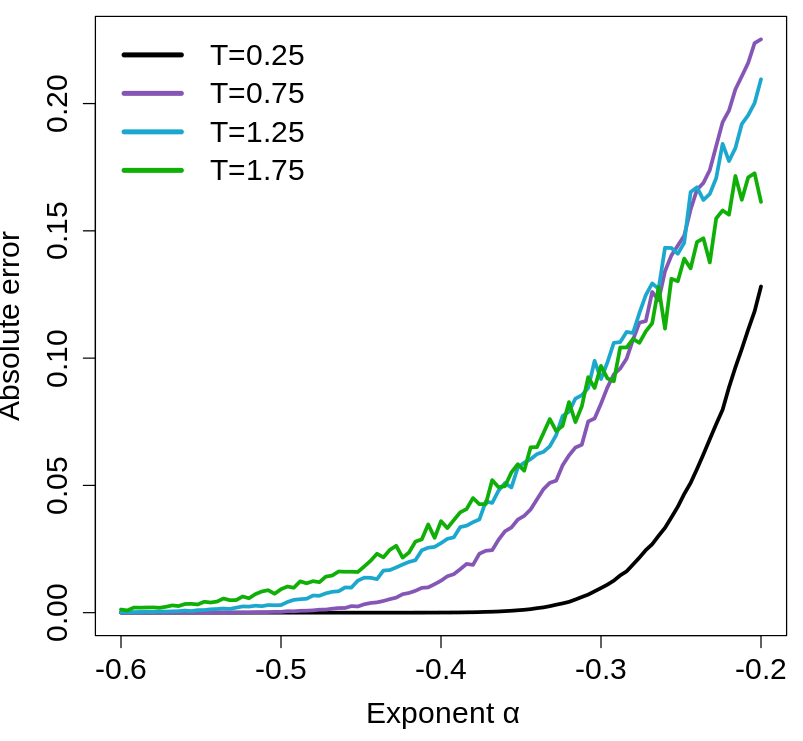}
\caption{\label{fig:alpha}\textbf{Absolute error of the WKB approximation as function of the exponent $\alpha$ and the maturity time}. For four values of the maturity time $T$, we computed the absolute error between the simulated option price and the WKB approximation. We observe that the error increases as the maturity time grows. In addition, exponents $\alpha$ below $-0.4$ (associated to moderate volatility skew) lead to negligible errors in option prices. Averages were taken over $10^6$ simulations (observe that the variability of the average increases for large maturities). Remaining model parameters for simulations are: initial stock price $S_0=100$, strike price $E=110$, interest rate $r=0.03$, drift constant $\mu=0.03$, and $\sigma=0.3$ (which is the prefactor in the diffusion term of Eq.~\eqref{eq:randCEV}, $\sigma S^{\alpha+1}$).} 
\end{figure}

Figure~\ref{fig:alpha} shows that the absolute error increases with increasing maturity time and when the exponent $\alpha$ (which controls the volatility skew, for $-1\le \alpha <0$) gets close to zero. Note that for $\alpha$ close to zero, volatility skew tends to zero as well (consistently with the fact that $\alpha\to 0$ reproduces the geometric Brownian motion assumed in the BS model). In fact, from Eq.~\eqref{eq:randCEV}, we observe that the volatility is equal to $\sigma S^{\alpha}$, and when $\alpha$ tends to $-1$, the volatility skew increases, while if $\alpha$ tends to zero, volatility becomes constant. Surprisingly, we find that the WKB approximation performs very well in extreme market conditions, with a high volatility skew and tail risk. As expected, the approximation works accurately for low maturities. We observe a non-monotonic effect when both $\alpha$ and $T$ increase: although increasing $\alpha$ augments the error, for a fixed exponent value the average error can decrease for larger maturity times. Therefore, a compensation of the error obtained for small $\alpha$ can occur if maturity times are large.

\begin{figure}[t!]
\centering
\includegraphics[width=\textwidth]{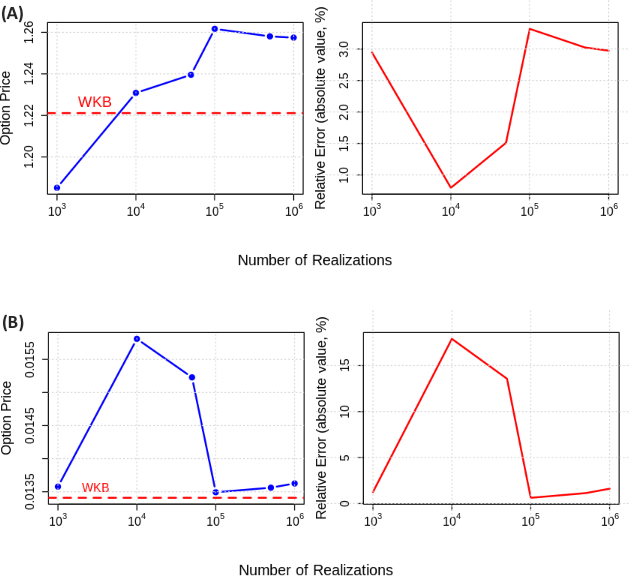}
\caption{\label{fig:relative}\textbf{Relative error of the WKB approximation as function of the number of Monte Carlo realizations}. We have carried out a varying number of Monte Carlo CEV model simulations, and obtained the average option price from simulated stock paths. In left panels, we show with a continuous line the option price as function of the number of Monte Carlo realizations, and the dashed horizontal line marks the value predicted by the WKB method. Right panels show the relative error (in absolute value) as function of the number of simulations. In all cases we used as parameter values: initial stock price $S_0=100$, strike price $E=110$, interest rate $r=0.03$, maturity time $T=2$, drift constant $\mu=0.03$, and varied $\sigma$ and the exponent $\alpha$, leading to the two rows of panels: (A) $\sigma = 0.3$ and $\alpha = -0.4$; (B) $\sigma = 0.2$, $\alpha = -0.6$. Observe that the WKB approximation performs very well (even for maturity times as long as $T=2$, see Fig.~\ref{fig:alpha}) and the relative error tends to stabilize as the number of simulations increases. The relative error can be moderately high (as in B), particularly because in that case the option value tends to be close to zero on average.} 
\end{figure}

Figure~\ref{fig:relative} shows the convergence of the simulated option prices as the number of simulated stochastic paths increases for two different numerical scenarios in which we vary the volatility prefactor $\sigma$ and the exponent $\alpha$. In addition, we show how the absolute value of the relative error depends on the number of realizations. Observe that the WKB approximation performs very well (note the scale of both vertical axes), with the relative error stabilizing as the number of simulations increases. In some cases the relative error may appear moderately high, primarily because the option value is, on average, close to zero (see panel B).

To conclude this Appendix, we provide in Table~\ref{tab:table} a detailed comparison of the performance of the semiclassical approximation for the CEV model. We set a reference value for relevant parameters, and compared the accuracy of the semiclassical approximation for the CEV model against averages over $10^6$ simulations of the option price, by varying each parameter separately while keeping the remaining ones constant. The results show that the absolute error yielded by the semiclassical approximation is small, except for values of $\alpha$ close to zero, a result consistent with the trends reported in Fig.~\ref{fig:alpha}, which can be associated to the singularity introduced in the model while changing variables from $S_t$ to $X_t$ ($S_t^{-2\alpha}=\sigma^2\alpha^2X_t$). Overall, this numerical results highlight the degree of accuracy of the semiclassical approximation, which reinforces (with another example) the goodness of the approximation observed in many contexts.

\begin{table}[h!]
\centering
\setlength\tabcolsep{5pt}
\begin{tabular}{|cc|cc|cc|cc|}
\hline
\multicolumn{2}{|c|}{$\bm{\alpha}$} &
\multicolumn{2}{|c|}{$\bm{T}$} &
\multicolumn{2}{|c|}{$\bm{\sigma}$} &
\multicolumn{2}{|c|}{$\bm{\mu}$} \\
\hline
Value & Absolute error &
Value & Absolute error &
Value & Absolute error &
Value & Absolute error \\
\hline
-0.9 & \(1.5241 \times 10^{-48}\) &
0.500 & \(1.5962 \times 10^{-5}\) &
0.0500  & \(1.3921 \times 10^{-42}\) &
0.010 & \(2.7923 \times 10^{-4}\) \\
-0.8 & \(6.9853 \times 10^{-21}\) &
0.625 & \(1.0219 \times 10^{-4}\) &
0.1125  & \(2.2606 \times 10^{-10}\) &
0.015 & \(4.6876 \times 10^{-4}\) \\
-0.7 & \(1.2885 \times 10^{-9}\) &
0.750 & \(3.2585 \times 10^{-4}\) &
0.1750  & \(6.7383 \times 10^{-6}\) &
0.020 & \(9.1751 \times 10^{-4}\) \\
-0.6 & \(2.2665 \times 10^{-5}\) &
0.875 & \(1.1808 \times 10^{-3}\) &
0.2375  & \(1.8603 \times 10^{-4}\) &
0.025 & \(1.3720 \times 10^{-3}\) \\
-0.5 & \(1.8319 \times 10^{-3}\) &
1.000 & \(1.9668 \times 10^{-3}\) &
0.3000  & \(1.5703 \times 10^{-3}\) &
0.030 & \(2.0402 \times 10^{-3}\) \\
-0.4 & \(2.1165 \times 10^{-2}\) &
1.125 & \(3.2811 \times 10^{-3}\) &
0.3625  & \(7.2910 \times 10^{-3}\) &
0.035 & \(2.3132 \times 10^{-3}\) \\
-0.3 & \(9.6261 \times 10^{-2}\) &
1.250 & \(4.2473 \times 10^{-3}\) &
0.4250  & \(1.7807 \times 10^{-2}\) &
0.040 & \(3.6850 \times 10^{-3}\) \\
-0.2 & \(2.1308 \times 10^{-1}\) &
1.375 & \(5.4281 \times 10^{-3}\) &
0.4875  & \(3.7453 \times 10^{-2}\) &
0.045 & \(4.7365 \times 10^{-3}\) \\
-0.1 & \(2.4028 \times 10^{-1}\) &
1.500 & \(7.0258 \times 10^{-3}\) &
0.5500  & \(6.4019 \times 10^{-2}\) &
0.050 & \(6.9212 \times 10^{-3}\) \\
\hline
\end{tabular}
\caption{Numerical evaluation of the performance of the semiclassical approximation for the CEV model. For parameters $(\alpha,T,\sigma,\mu)$ we defined a reference set of values, i.e., $(\alpha^{\star},T^{\star},\sigma^{\star},\mu^{\star})=(-0.5,1,0.3,0.03)$ and varied each parameter (``Value'' columns) while keeping the remaining ones constant and equal to the starred values. The initial stock price was set to $S_0=100$ and the strike price to $E=110$. Then we computed the absolute error between averages of the price option over $10^6$ realizations and the estimate provided by the semiclassical approximation (``Absolute error'' columns). The reference values were chosen as typical parameters of the model (usual maturity times and volatility). To keep neutral-risk assumptions, we took $\mu=r\approx 0.03$, a typical interest rate, and we did not varied the interest rate $r$ given the neutrality assumption. The results evidence the accuracy of the approximation, with absolute errors less than $\sim 5\times 10^{-2}$, except for values of $\alpha$ close to zero, for which the approximation performs worse.
}
\label{tab:table}
\end{table}

\end{document}